\documentclass[pra,twocolumn,aps,superscriptaddress,showpacs]{revtex4}
\usepackage{epsfig}
\usepackage{graphicx}
\usepackage{amsmath}
\usepackage{amscd}

\newcommand{\beq}{\begin{eqnarray}}
\newcommand{\eeq}{\end{eqnarray}}
\def\be{\begin{equation}}
\def\ee{\end{equation}}
\def\ba{\begin{eqnarray}}
\def\ea{\end{eqnarray}}

\begin{document}

\title
{Synthetic Spin-Orbit Coupling in Two-level Cold Atoms }

\author{Qi Zhang}
\affiliation{Laboratory of Condensed Matter Theory and Computational Materials, Zhengzhou University,
Zhengzhou 450052, People's Republic of China}
\affiliation{Center for Clean Energy and Quantum Structures and School of Physics and Engineering,
Zhengzhou University, Zhengzhou 450052, People's Republic of China}
\affiliation{Centre of Quantum Technologies and Department of
Physics, National University of Singapore, 117543, Singapore}

\author{Jiangbin Gong}
\affiliation{Department of Physics and Centre for Computational
Science and Engineering, National University of Singapore, 117542,
Singapore} \affiliation{NUS Graduate School for Integrative Sciences
and Engineering, Singapore 117597, Singapore}

\author{C.H. Oh}
\affiliation{Centre of Quantum Technologies and Department of
Physics, National University of Singapore, 117543, Singapore}
\affiliation{Institute of Advanced Studies, Nanyang Technological
University, Singapore 639798, Singapore}

\date{\today}
\begin{abstract}
%Zitterbewegung (ZB, trembling motion in free space) occurs in
%particles with spin-orbit coupling (SOC), which, for cold atoms, is traditionally
%implemented by ``tripod" scheme with four
%-level atoms interacting with three
%laser fields.
Synthetic spin-orbit coupling (SOC) in controlled quantum systems such as cold atoms or trapped ions has been
of great interest.  Here we show, both theoretically and computationally, a simplest realization
of SOC using two-level cold atoms interacting with only one laser beam. The underlying mechanism
is based upon the non-adiabatic nature of laser-atom interaction, with
the Rabi frequency and atom's kinetic energy being comparable to each other.
We use the Zitterbewegung (ZB) oscillation to further illustrate the effects of the
synthesized SOC on the quantum dynamics of the two-level cold atoms.
We expect our proposal to be of experimental interest in quantum simulation of SOC-related physics.
\end{abstract}
 \maketitle

\section{introduction}
Artificial or synthetic SOC in a number of quantum systems has been realized.
Examples include trapped-ion systems \cite{LamataPRL,GerritamaNature}, band
electrons in graphene \cite{Novoselov}, cavity electrodynamics
\cite{larson}, macroscopic sonic crystals and photonic superlattices \cite{ZhangPRL2008}, as well as the more controllable ultracold atoms in designed laser fields \cite{JuzeliunasPRA2008,VaishnavPRL2008,MerklEPL2008,Wang2010PRL}. In particular,
one intensively studied model for implementing SOC in  cold atoms is the so-called
``tripod" scheme, where three levels with different magnetic quantum numbers are coupled with a common excited state by
three laser beams \cite{JuzeliunasPRA2008,VaishnavPRL2008,MerklEPL2008,Wang2010PRL}. Therein
two degenerate dark states are formed, and due to the the center of mass motion of the cold atoms,
the two dark states are coupled together and as a result an effective SOC emerges.
However, because the three levels on the ground state manifold with different magnetic quantum numbers are perturbed by
magnetic field fluctuations differently, this ``tripod" scheme faces the challenging issue of dephasing and to our knowledge,
it is yet to be realized in an actual cold-atom experiment.  As an exciting progress,
cold-atom SOC was realized in Ref. \cite{Lin2011Nature} using a different and somewhat complicated scheme, where
two of three atomic internal states are dressed by a pair of Raman laser beams in the presence of a biased magnetic field.

In this paper we propose the simplest scheme to date for the realization of cold-atom SOC, where a two-level
atom interacts with one laser beam only.  In many previous proposals for realizing an artificial vector gauge potential acting on neutral atoms, the Rabi frequency (optical strength) is set to be much larger than other parameters of the system (for instance, the atomic kinetic energy), such that the optical field is strong enough to determine the atomic internal state (dressed state) during the center of mass motion of the cold atom. In other words, the internal state of the cold-atom is assumed to be adiabatically following the center of mass motion. By contrast, here we set the Rabi frequency at a level comparable to cold atom's kinetic energy.  This leads to a non-adiabatic coupling between the center of mass motion and the internal states.  Naturally, the non-adiabatic coupling depends sensitively on the moving speed of the atom and thus creates a type of SOC \cite{gongprl2002} (In Ref.~\cite{gongprl2002}, the non-adiabatic transitions or the resulting SOC can even induce chaos in the center of mass motion). As shown below, the explicit form of the coupling in our two-level setup indicates that
a simple version of SOC can be obtained.

To elaborate on the dynamical effects of the synthetic SOC in our setup, we consider the jittering center of mass motion of the cold-atom
as an analog of the
Zitterbewegung (ZB) \cite{Schrodinger} of electrons described by Dirac's equation. This is of experimental interest
because the true ZB associated
with a real Dirac particle (e.g., electron and neutrino) cannot be
observed due to its extremely large frequency and small amplitude
(it does not exist even when considering the quantum field theory
\cite{Krekora}).  Note that artificial ZB has been experimentally implemented in trapped
ions \cite{GerritamaNature}, which requires several beams of lasers, whereas in our case
only one laser beam will be needed. This simplification creates more freedom for further quantum control purposes.

This short report is organized as follows.
In Section 2 we shall derive synthetic SOC in a two-level cold atom system,
of which the two levels are coupled by one plane-wave laser field. In Section 3 we derive the ZB oscillation due to
SOC. To confirm our theoretical considerations and
 to illustrate that the SOC is nontrivial,
we also show computational results of ZB.  Section 4 concludes this work.

\section{Simple Theory}

Consider a two-level atom interacting with a plane wave laser beam. The total Hamiltonian reads,
\begin{equation}\label{originalH}
    \hat{H}=\frac{-\hbar^2\nabla^2}{2m}+\left(\begin{array}{cc}\epsilon/2&2\hbar\Omega_0\cos(kz+\omega t)\\ 2\hbar\Omega_0\cos(kz+\omega t)&-\epsilon/2\end{array}\right),
\end{equation}
where $\frac{-\hbar^2\nabla^2}{2m}$ is atom's kinetic energy term, $m$ being the atom mass; $\epsilon$ is
the energy bias between the two internal levels; $k$ and $\omega$ are the wave vector and frequency of the
 laser field; $\Omega_0$ defines the Rabi frequency of this laser-atom system. One normally assumes
 $\omega=ck$, with $c$ being the speed of light in free space. However, one may also
  modify this relation by tilting the laser beam. Without loss of generality, we assume that the laser beam travels in $z$ direction. The total wavefunction can be denoted as $|\psi\rangle=\phi(\mathbf{r})(a,b)^T$, with $\phi(\mathbf{r})$ and $(a,b)^T$ respectively associated with the spatial and internal parts of the total wavefunction. The total wavefunction satisfies the Schr\"odinger equation $i\hbar\frac{\partial}{\partial t}|\psi\rangle=\hat{H}|\psi\rangle$.
It should be noted here that
analogous laser-atom Hamiltonian (\ref{originalH}) was studied in the past. For example, it was used to study the energy bands of cold atoms in optical lattice (e.g., band structure, spontaneous emission of a two-level atom, atom's effective mass) \cite{past}. Our emphasis here is placed on the plane-wave states of the cold atom. Physically, a wavepacket with sufficient width in the coordinate space will be describable by a plane wave.

We assume that the energy bias $\epsilon$ between the two internal levels is much larger than the Rabi frequency and that
 the laser frequency is on resonance with the atomic transition. We can then exploit the well-known rotating wave approximation (RWA) (the validity of this approximation will be checked in our computational studies below). That is, under the resonance condition $\omega=\epsilon$, we make a wavefunction transformation from $|\psi\rangle$ to $|\psi'\rangle=\phi(\mathbf{r})(a',b')^T$,
\begin{equation} \label{transformation1}
 a'=a\exp(i\omega t/2), b'=b\exp(-i\omega t/2).
\end{equation}
In the $(a',b')$ representation,  the Hamiltonian (after dropping the counter-rotating terms) becomes
\begin{equation}\label{RWAH}
    \hat{H}_{\text{RWA}}=\frac{-\hbar^2\nabla^2}{2m}+\left(\begin{array}{cc}0&\hbar\Omega_0\exp(-ikz)\\ \hbar\Omega_0\exp(ikz)&0\end{array}\right),
\end{equation}
We remark that the above representation transformation (\ref{transformation1}) is independent of the spatial coordinate, so the kinetic energy term retains its original form.

The time-independent RWA Hamiltonian in Eq.~(\ref{RWAH}) depends on both momentum operator $-i\hbar\nabla$ and coordinate $z$. An important
 observation is  that the Hamiltonian depends on $z$ through special factors, i.e., $\exp(\pm i kz)$.  This observation leads to the expectation that the Hamiltonian eigenstates should  assume a plane wave form. This is indeed the case and can be seen more explicitly after making another unitary transformation from $|\psi'\rangle$ to $|\psi''\rangle=\phi(\mathbf{r})(a'',b'')^T$, with
\begin{equation} \label{transformation2}
a''=a'\exp(ikz/2), b''=b'\exp(-ikz/2).
\end{equation}
In the $(a'',b'')$ representation, one can prove from the Schr\"odinger equation that, if the spatial part of the wavefunction assumes a plane-wave form $\phi(\mathbf{r})=(2\pi\hbar)^{-3/2}e^{\frac{i}{\hbar}(p_xx+p_yy+p_zz)}$, the effective Hamiltonian can be written in a coordinate-independent form,
\begin{equation}\label{finalH1}
    \hat{H}_{\text{SOC}}=\left(\begin{array}{cc}-\frac{\hbar k}{2m}p_z&\hbar\Omega_0\\ \hbar\Omega_0&\frac{\hbar k}{2m}p_z\end{array}\right)+\frac{p_x^2+p_y^2+p_z^2}{2m}+\frac{\hbar^2k^2}{8m},
\end{equation}
which already shows the emergence of SOC (i.e., SOC in the form of $\sigma_z p_z$, where $\sigma_z$ is a Pauli matrix).
As seen above, the energy splitting in the $(a'',b'')^T$ representation depends on momentum $p_z$.  %The previous assumption that an eigenstate should possess a plane-wave form is also verified.
 It is also seen that the momenta in $x$ and $y$ directions do not contribute to SOC and that
 the term $\frac{\hbar^2k^2}{8m}$ is merely a constant. For this reason we take $p_x=p_y=0$ for convenience
 and drop the constant.  The final one dimensional (1D) effective Hamiltonian becomes
\begin{equation}\label{finalH}
    \hat{H}_{\text{SOC}}^{\text{1D}}=\left(\begin{array}{cc}-\frac{\hbar k}{2m}p_z&\hbar\Omega_0\\ \hbar\Omega_0&\frac{\hbar k}{2m}p_z\end{array}\right)+\frac{p_z^2}{2m}.
\end{equation}

Thus with only one laser beam, 1D SOC (SOC associated with one component of momentum) in two-level atoms can be
 realized.  Though the well-known ``tripod" scheme using four-level atoms and three laser beams can realize 2D SOC (SOC associated with two components of momentum, e.g., Rashba or Dresselhaus type) instead of 1D SOC,
  we note that 1D SOC is sufficient for many interesting purposes and sometimes 1D SOC is even more advantageous.
 We also emphasize that in previous approaches to synthetic gauge fields, the Rabi frequency is always taken to be much larger than atom's kinetic energy \cite{Ruseckas2005PRL,JuzeliunasPRA2008}; whereas here the Rabi frequency $\Omega_0$ should be comparable to atom's kinetic energy.

 It is also interesting to compare
 our proposal with the realized mechanism in the experiment in Ref.~(\cite{Lin2011Nature}). SOC in both cases can be interpreted as an outcome
 of the non-commutability between local spin rotation and the center of mass motion of the cold atom.  The two-level set up proposed here is however
 more straightforward. The main assumption made here is that the lifetime of the atom's bare excited state is long enough such that the spontaneous decay is negligible during a time scale of interest to observe the SOC effect. Fortunately, this assumption can be fulfilled in real experiments. For example, ytterbium and alkaline-earth atoms have a spin-singlet ground state and an extremely long-lived spin triplet excited state. Indeed, the lifetime of the spin triplet excited state for Yb is almost around $20$ s \cite{PorsevPRA}.

\section{ZB oscillations}
ZB oscillations are now one of the most known effects of SOC \cite{JuzeliunasPRA2008,VaishnavPRL2008,MerklEPL2008,ZhangPRAzb,ZhangEPL,GerritamaNature}.
Interestingly, in the case of 1D SOC, the damping of ZB can be much slower than that in the 2D SOC case \cite{MerklEPL2008,ZhangPRAzb,ZhangEPL,GerritamaNature}. This difference has been explained by analyzing the group velocity of time evolving wavepackets undergoing ZB oscillations\cite{ZhangPRAzb}.
Following the very same procedure as in Refs. \cite{MerklEPL2008,ZhangPRAzb,ZhangEPL},
we first take an initial wavepacket in $(a'',b'')$ representation as (taking $p_z=0$ for simplicity),
\begin{equation} \label{inistate}
|\psi''_{\text{ini}}\rangle=g(z)\frac{1}{\sqrt{2}}\left(\begin{array}{c}1\\-i\end{array}\right),
\end{equation}
with $g(z)$ being a Gaussian center on $z=0$. The average position of the subsequent matter wave can be analytically derived as
\begin{eqnarray} \label{ZB} \nonumber
\langle z\rangle&=&\langle\psi''_{\text{ini}}|e^{i\hat{H}_{\text{SOC}}^{\text{1D}}t}ze^{-i\hat{H}_{\text{SOC}}^{\text{1D}}t}
|\psi''_{\text{ini}}\rangle \\
&=&\frac{\hbar k}{2m\Omega_0}\sin^2(\Omega_0t),
\end{eqnarray}
where $e^{-i\hat{H}_{\text{SOC}}^{\text{1D}}t}$ is the time evolution operator.

In deriving Eq.~(\ref{ZB}), we have made the wide-wavepacket approximation as in Ref. \cite{VaishnavPRL2008,MerklEPL2008,ZhangPRAzb,ZhangEPL}, i.e., the width of Gaussian $g(z)$ in coordinate space $\delta z$ is large enough such that the width of the corresponding wavepacket in momentum space $\delta p_z (\sim\hbar/\delta z)$ satisfies
\begin{equation} \label{zzz}
\frac{k\delta p_z}{m}\ll\Omega_0.
\end{equation}
The consequence of the wide-wavepacket approximation is twofold. First, according to the Hamiltonian (\ref{finalH}) the spinor will ``rotate" around the effective ``magnetic field" proportional to $\hbar\Omega_0\hat{x}-\frac{\hbar k}{2m}p_z\hat{z}$. The wavepacket in momentum space $\delta p_z$ is very small, so we can take the approximation that all the effective ``magnetic fields" are the same in magnitude (but they are different in directions). As a result, the same initial spinor $(1,-i)^T/\sqrt{2}$ rotates about different axis with the same frequency \cite{VaishnavPRL2008,ZhangEPL}. The evolution for the spatial degree of freedom and that for the internal state are thus coupled together. Second, according to Eq.~(\ref{zzz}), the wide-wavepacket approximation determines that the amplitude of ZB is much smaller than the width of the atomic wavepacket
\begin{equation}
\frac{\hbar k}{2m\Omega_0}\ll\delta z.
\end{equation}
The large ZB amplitude can only arise when $\delta p_z$ becomes larger, or $\frac{k\delta p_z}{m}\sim\Omega_0$. However, in that case the rotational frequencies of spinors are different for different magnitudes of ``magnetic field" and Eq. (\ref{ZB}) does not hold anymore, giving rise to non-negligible deforming and fast decaying of ZB \cite{VaishnavPRL2008}.

 The small ZB amplitudes are not because of the particular realization of ZB in our scheme:  it actually stems from the quantum nature of atom's center of mass motion \cite{VaishnavPRL2008,MerklEPL2008,ZhangPRAzb} (only if the center of mass motion is treated classically can the ZB amplitude tuned to be arbitrarily large).  To enhance the ZB oscillations, one may further introduce a driven term in SOC, where the amplitude of ZB can be increased or even converted to directed motion \cite{ZhangEPL}. According to Ref.~\cite{ZhangEPL}, we may modulate the Rabi frequency periodically to enhance the ZB oscillations. We do not pursue such details here.

We next discuss the $(a'',b'')$ representation used in our theory. This dressed-state representation is related to the original bare basis states by two unitary transformations, one connected with the optical frequency and the other one connected with the laser wave vector. Note that both transformations (\ref{transformation1}) and (\ref{transformation2}) do not affect the average position of the moving cold atom, so
the predicted ZB effect occurs in all the three representations.
In order to make clear the actual physical picture, we now rewrite the initial state (\ref{inistate}) in the original representation when $t=0$,
\begin{equation} \label{inistateO}
|\psi''_{\text{ini}}\rangle=g(z)\frac{1}{\sqrt{2}}\left(\begin{array}{c}\exp(-ikz/2)\\-i\exp(ikz/2)\end{array}\right),
\end{equation}
which marks the superposition of two wavepackets with opposite central wave-vectors. In a lab-frame free space, the two wavepackets possess opposite group velocities $\mp\hbar k/m$. As a result, without a laser field the initial atomic wavepacket will split into two parts. Interestingly, in the presence of the laser field, the initial state (\ref{inistateO}) does not split but undergoes ZB oscillation because the atomic Hamiltonian is modified by the laser field. The initial state preparation here is also simpler than that required by early proposals for observing ZB oscillations based on the ``tripod-scheme \cite{JuzeliunasPRL2008,ZhangPRA2009}.

Our theoretical prediction [Eq.~(\ref{ZB})] is also verified by fully numerical simulations starting from the very initial Hamiltonian (\ref{originalH}) without any  approximation. The well-known split-operator technique is used in our computational studies. One computational result is shown in Fig. 1, which clearly shows that (i) the ZB period is around $\pi$ in units of $1/\Omega_0$ (the corresponding frequency is $2$ in units of $\Omega_0$) and (ii) the ZB amplitude is around $0.5$ in units of $\hbar k/m\Omega_0$, in full agreement with the theoretical result of Eq.~(\ref{ZB}).

\begin{figure}[t]
\begin{center}
\vspace*{-0.5cm}
\par
\resizebox *{8cm}{6cm}{\includegraphics*{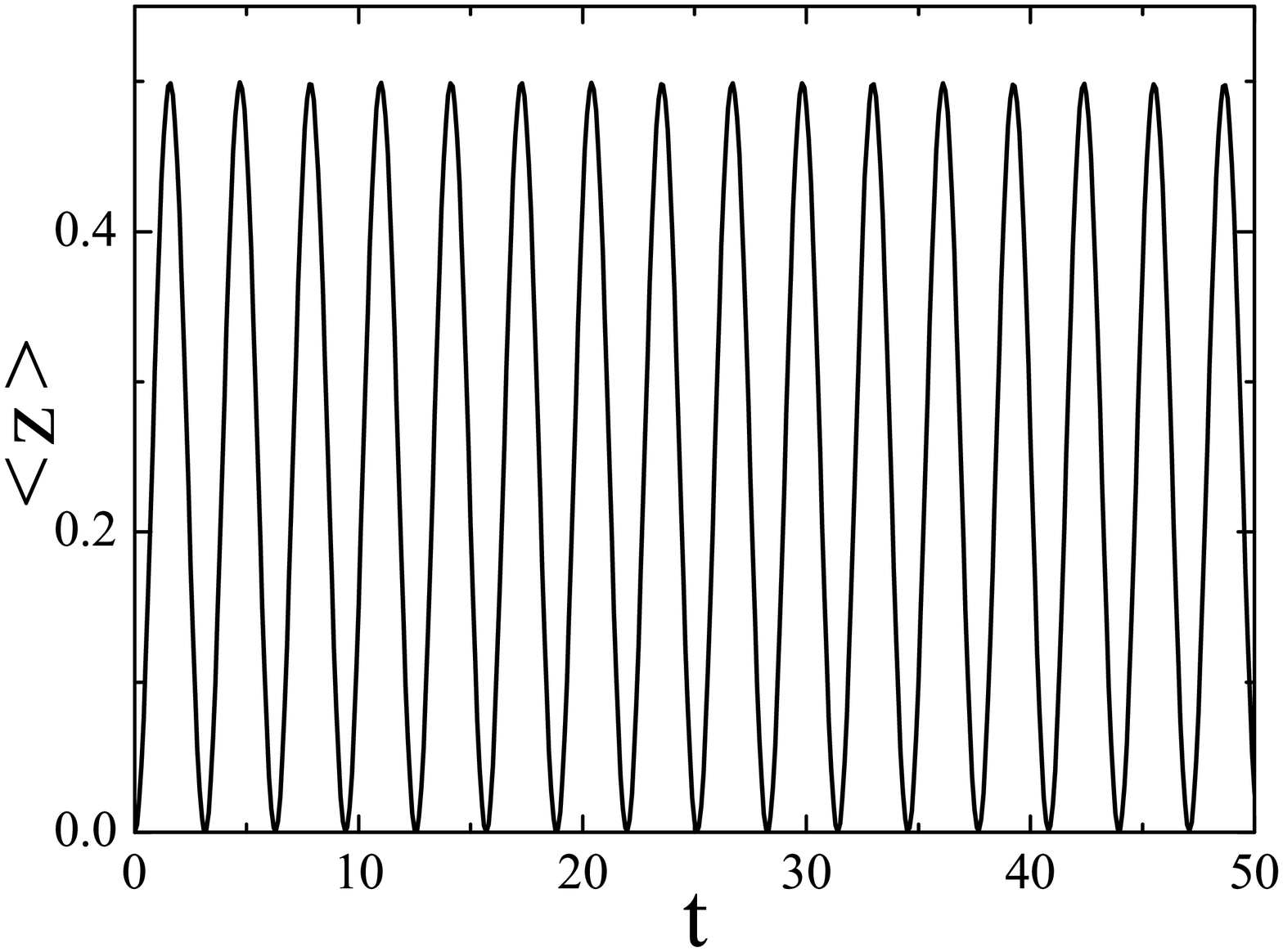}}
\end{center}
\par
\vspace*{-0.5cm} \caption{Numerical result of ZB oscillation with the initial state taken as (\ref{inistate}). The standard
split-operator technique is used to simulate the dynamics based on the original Hamiltonian (\ref{originalH}) without making any approximation,
with $\epsilon=\omega=50$ (in units of $\Omega_0$). In the simulation, the width of the Gaussian wavepacket is taken as $\delta z=30$, which almost keeps a constant during the shown time scale. Throughout $t$ is in units of $1/\Omega_0$, z is in units of $\hbar k/m\Omega_0$. The numerical results fully agree with our theoretical analysis.}\label{fig1}
\end{figure}

\section{Concluding Remarks}

To conclude, we have proposed a straightforward implementation of SOC using two-level atoms interacting with one laser beam.  As indicated by our analysis above, the SOC coupling strength is easily tunable because it is proportional to the effective wave vector of the laser beam along the direction of center of mass motion (effective wave vector changes when the laser beam is tilted).  Further, the magnitude of the $\sigma_x$ term, which can be regarded as
the ``mass" term of a Dirac particle [after neglecting the $p_z^2$ term in Eq.~(6)], can also be adjusted by varying the Rabi frequency, either spatially or temporally.

This work indicates that non-adiabatic coupling in laser-atom interaction can be very useful in quantum simulation.
Indeed, the central idea behind our proposal is to directly exploit  non-adiabatic effects in the coupling between the center of mass motion and the internal states of long-lived two-level atoms.
Our scheme is hoped to further simplify experimental studies of SOC in the cold-atom context.
We have also used the ZB oscillations to elaborate on how SOC affects the center of mass motion in our setup.
It is found that both the amplitude and frequency of ZB depend on the Rabi frequency, which suggests a more flexible control over ZB. If we set $m\sim10^{-25}\text{Kg}$, $k\sim10^6\text{m}^{-1}$, we find that a ZB amplitude $10^{-8}\text{m}$ is obtained if the Rabi frequency is taken as $\Omega_0\sim10^5\text{Hz}$.  A much larger ZB amplitude, say, $1$mm, is also possible if the Rabi frequency is decreased to $\Omega_0\sim10^0\text{Hz}$, with a ZB frequency about $10^0$Hz.  Both the simplicity and flexibility of our two-level-atom scheme should be of interest to ongoing studies of spin-orbit coupling
related physics.

\section{acknowledgments}

 This work was supported by the National Natural Science Foundation of China (Grant No. 11105123).

\end{document}